\documentclass[%
	conference,
	a4paper,
	10pt
]{IEEEtran}

\usepackage{graphicx}
\usepackage{epstopdf}

\usepackage{standalone}		
\usepackage[nolist ]{acronym}
\usepackage{xstring}

\usepackage[utf8]{inputenc}
\usepackage{marvosym}

\usepackage{algorithm}
\usepackage{algpseudocode}
\usepackage{tikz}

\usepackage{pgfplots}
\pgfplotsset{compat=newest}
\pgfplotsset{
    box plot/.style={
        /pgfplots/.cd,
        fill=blue!30,
        only marks,
        mark=-,
        mark size=0.2em,
        /pgfplots/error bars/.cd,
        y dir=plus,
        y explicit,
    },
    box plot box/.style={
        /pgfplots/error bars/draw error bar/.code 2 args={%
            \draw  ##1 -- ++(.2em,0pt) |- ##2 -- ++(-.2em,0pt) |- ##1 -- cycle;
        },
        /pgfplots/table/.cd,
        y index=2,
        y error expr={\thisrowno{3}-\thisrowno{2}},
        /pgfplots/box plot
    },
    box plot top whisker/.style={
        /pgfplots/error bars/draw error bar/.code 2 args={%
            \pgfkeysgetvalue{/pgfplots/error bars/error mark}%
            {\pgfplotserrorbarsmark}%
            \pgfkeysgetvalue{/pgfplots/error bars/error mark options}%
            {\pgfplotserrorbarsmarkopts}%
            \path ##1 -- ##2;
        },
        /pgfplots/table/.cd,
        y index=4,
        y error expr={\thisrowno{2}-\thisrowno{4}},
        /pgfplots/box plot
    },
    box plot bottom whisker/.style={
        /pgfplots/error bars/draw error bar/.code 2 args={%
            \pgfkeysgetvalue{/pgfplots/error bars/error mark}%
            {\pgfplotserrorbarsmark}%
            \pgfkeysgetvalue{/pgfplots/error bars/error mark options}%
            {\pgfplotserrorbarsmarkopts}%
            \path ##1 -- ##2;
        },
        /pgfplots/table/.cd,
        y index=5,
        y error expr={\thisrowno{3}-\thisrowno{5}},
        /pgfplots/box plot
    },
    box plot median/.style={
        /pgfplots/box plot
    },
    boxplot/every median/.style={
    	ultra thick,dashed,cyan
    }
}

\definecolor{flexicolor}{RGB}{46,49,146}
\definecolor{amaricolor}{RGB}{237,28,36}

\usepackage{xspace}

\usepackage[binary-units=true]{siunitx}
\usepackage{psfrag}
\usepackage{graphicx}
\usepackage{tabularx,booktabs}
\usepackage{multirow}
\usepackage{rotating}

\renewcommand{\baselinestretch}{1}

\usepackage{multirow}

\usepackage[cmex10]{amsmath}
\usepackage[caption=false,font=footnotesize]{subfig}
\usepackage{stfloats}
\renewcommand{\baselinestretch}{0.981}

\makeatletter
\renewcommand{\ALG@beginalgorithmic}{\small}
\makeatother

\begin{document}

\begin{acronym}
	\acro{PDR}{packet delivery ratio}
	\acro{CBR}{constant bitrate}
	\acro{CSV}{comma separated values}
	\acro{MA-OLSR}{mobility-aware OLSR}
	\acro{OLSR}{Optimized Link State Routing}
	\acro{B.A.T.M.A.N.}{Better Approach To Mobile Ad-hoc Networking}
	\acro{AODV}{Ad Hoc On-Demand Distance Vector}
	\acro{PASER}{Position-Aware, Secure, and Efficient Routing}
	\acro{HWMP}{Hybrid Wireless Mesh Protocol}
	\acro{ZRP}{Zone Routing Protocol}
	\acro{GPSR}{Greey perimeter stateless routing}
	\acro{RGR}{Reactive-greedy-reactive}
	\acro{MANET}{mobile ad-hoc network}
	\acro{VANET}{vehicular ad-hoc network}
	\acro{UDP}{User Datagram Protocol}
	\acro{TC}{Topology Control}
	\acro{OMNeT++}{Objective Modular Network Testbed in C++}
	\acro{MAC}{Medium Access Control}
	\acro{GNSS}{global navigation satellite system}
	\acro{UAV}{Unmanned Aerial Vehicle}
	\acro{UGV}{Unmanned Ground Vehicle}
\end{acronym}

\acresetall

%
\title{An OMNeT++ based Framework for Mobility-aware Routing in Mobile Robotic Networks}

\author{\IEEEauthorblockN{\textbf{Benjamin Sliwa, Christoph Ide and Christian Wietfeld}}
\IEEEauthorblockA{Communication Networks Institute\\
TU Dortmund University\\
44227 Dortmund, Germany\\ e-mail: $\{$Benjamin.Sliwa, Christoph.Ide, Christian.Wietfeld$\}$@tu-dortmund.de}}


\maketitle

\begin{abstract}
In this paper, we propose a cross-layer extension for the INETMANET framework of OMNeT++, which utilizes mobility control knowledge to enhance the forwarding of routing messages. The well-known mobility meta-model from Reynolds is used to provide a realistic representation of the mobility behavior of autonomous agents with respect to the various influences those agents have to face in real-world applications.
Knowledge about the current mobility properties and its predicted development is used by mobility-aware routing mechanisms in order to optimize routing decisions and avoid path losses at runtime.
In a proof of concept evaluation we show that our proposed methods can achieve significant improvements to the robustness of communication paths. 
\end{abstract}

\IEEEpeerreviewmaketitle

\section{Introduction}
\ac{OMNeT++}~\cite{VargaHornig2008} is a well-established network simulation framework and has been extended with many \ac{MANET} routing protocols by the INETMANET framework. Although mobile nodes are supported and several generic mobility models are available, geo-assisted routing protocols are not yet part of the framework. Furthermore, the provided mobility models are not capable of providing a realistic represention of the high number of challenges mobile robotic systems have to face in real-world applications. Besides exploration tasks, collision avoidance and maintenance of the swarm coherence are further important aspects, which influence the mobility behavior.
In our recent work \cite{SliwaBehnkeIdeEtAl2016}, we presented the mobility-aware routing protocol B.A.T.Mobile and its trajectory prediction method, which leverages application layer mobility control knowledge for precise position estimations. We demonstrated its capability to provide reliable communication even under highly challenging channel conditions and showed it is significantly outperforming the established protocols in highly dynamic networks.
In this paper, we propose an extension of the capabilities of OMNeT++/INETMANET. Our goal is to provide a realistic mobility model for autonomous agents and to offer a solid basis for the design of further cross-layer and mobility-aware routing protocols. Therefore, we abstract the principles used by B.A.T.Mobile and propose a base module for protocols, which utilize knowledge about the controlled mobility trajectories to optimize the routing decisions. 
The remainder of this paper is structured as follows: After discussing the related work, we present the system model of our framework and discuss the relevant submodules. In the next section, we describe the simulation setup for a defined reference scenario. Finally, an impression about the potentials of the proposed approach is given by presenting results for a proof of concept evaluation.



\section{Related Work}

\begin{figure}[b!]  	
\vspace{-10pt}
	\centering		  
	\includegraphics[width=1\columnwidth]{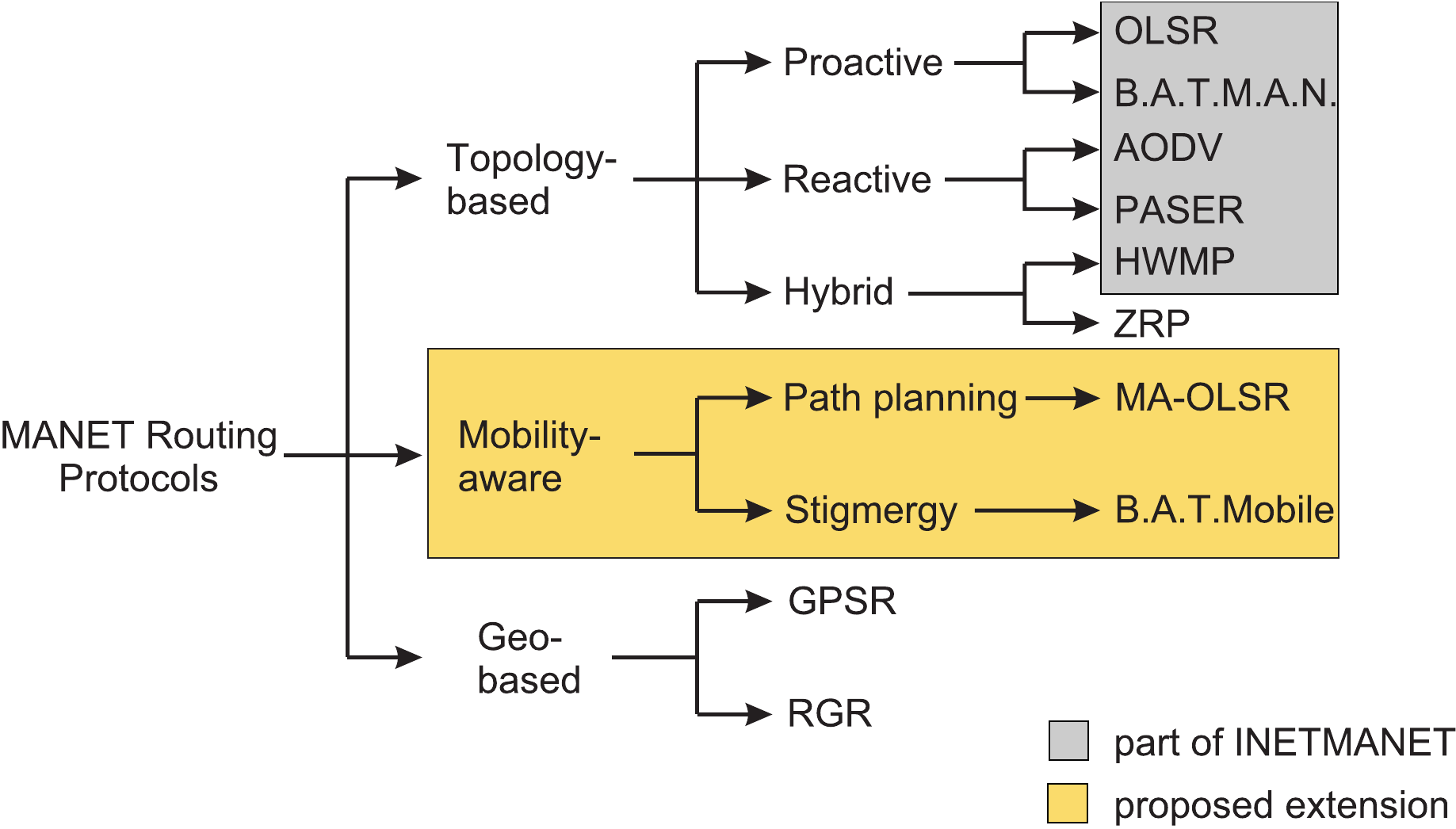}
	\caption{Classification of MANET routing protocols}
	\label{fig:protocols}
\end{figure}

In the field of \acp{MANET} various protocols have been proposed to face the various challenges mobile nodes encounter in real-world applications.
Fig.~\ref{fig:protocols} illustrates the different categories of routing approaches and names example protocols.
\emph{Topology-based} protocols use knowledge about the links between nodes for their decision-making and form the most popular category of \ac{MANET} routing approaches. Further distinction is done into proactive \cite{ClausenJacquet2003}\cite{JohnsonNtlatlapaAichele2008}, reactive \cite{PerkinsBelding-RoyerDas2003}\cite{Sbeiti2015} and hybrid \cite{Bahr2006}\cite{HaasPearlmanSamar2002} protocols.
Recently, the \emph{geo-based} approach \cite{KarpKung2000}\cite{Shirani2011} has received a lot of attention in the scientific community. Nodes determine their own position using a \ac{GNSS} receiver and manage knowledge about the positions of the other network participants with a \emph{location service}. The selection of forwarder nodes is performed depending on their distance to the destination. Different forwarding schemes are compared in \cite{MauveWidmerHartenstein2001}. Since link-state information is not considered by most geo-based protocols, additional recovery strategies are needed if the location-based topology map mismatches the actual channel conditions.
In this paper, we combine the \emph{topology-based} and the \emph{geo-based} approach to take advantage of both paradigms. The \emph{mobility-aware} routing class uses link-state information as well as knowledge about the mobility of the nodes to enhance the routing process. In earlier work we presented the B.A.T.Mobile protocol, which uses stigmergic principles. Later in this paper, we will describe a mobility-aware path planning algorithm and give a basic evaluation of the method as an extension to OLSR, called \ac{MA-OLSR}.

\section{Design and Implementation of the Framework}
\begin{figure}[ht!]  
	\vspace{-5pt}	
	\centering		  
	\includegraphics[width=1\columnwidth]{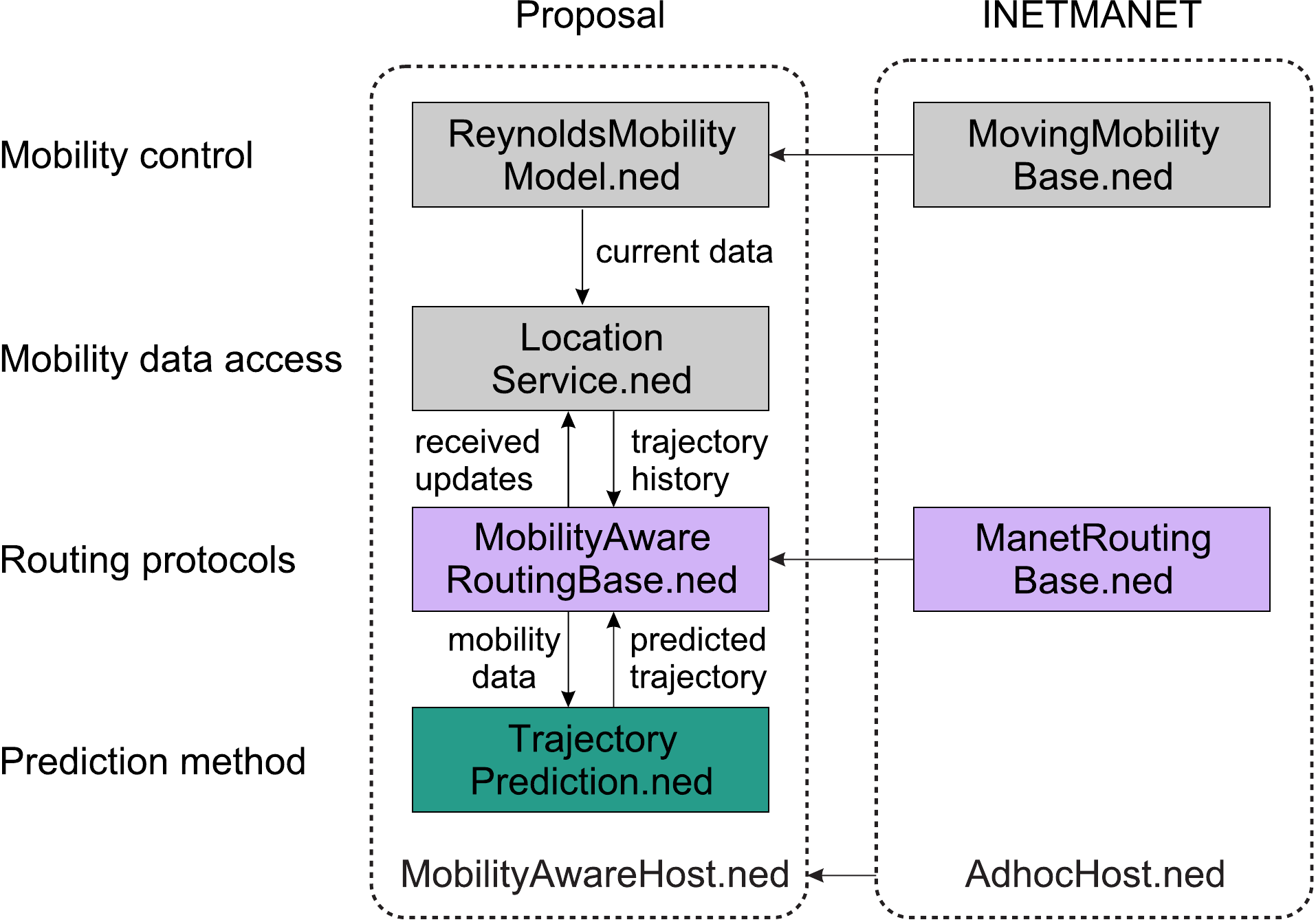}
	\caption{Cross-layer approach: utilization of mobility control layer information to enhance the routing process}
	\label{fig:approach}
	\vspace{-5pt}
\end{figure}

The proposed cross-layer system model is illustrated in Fig.~\ref{fig:approach}. The modules' implementations are derived by inheriting from the base modules of the INETMANET framework. The novel \emph{MobilityAwareHost.ned} acts as a compound module for all logical submodules required for mobility-aware routing.

In the following sub-chapters, we give a detailled description of the implementation of the mobility meta-model as well as the routing principles.

\subsection{Implementation of a realistic mobility meta-model for autonomous agents}

\begin{figure}[ht!]  	
	\centering		  
	\includegraphics[width=0.8\columnwidth]{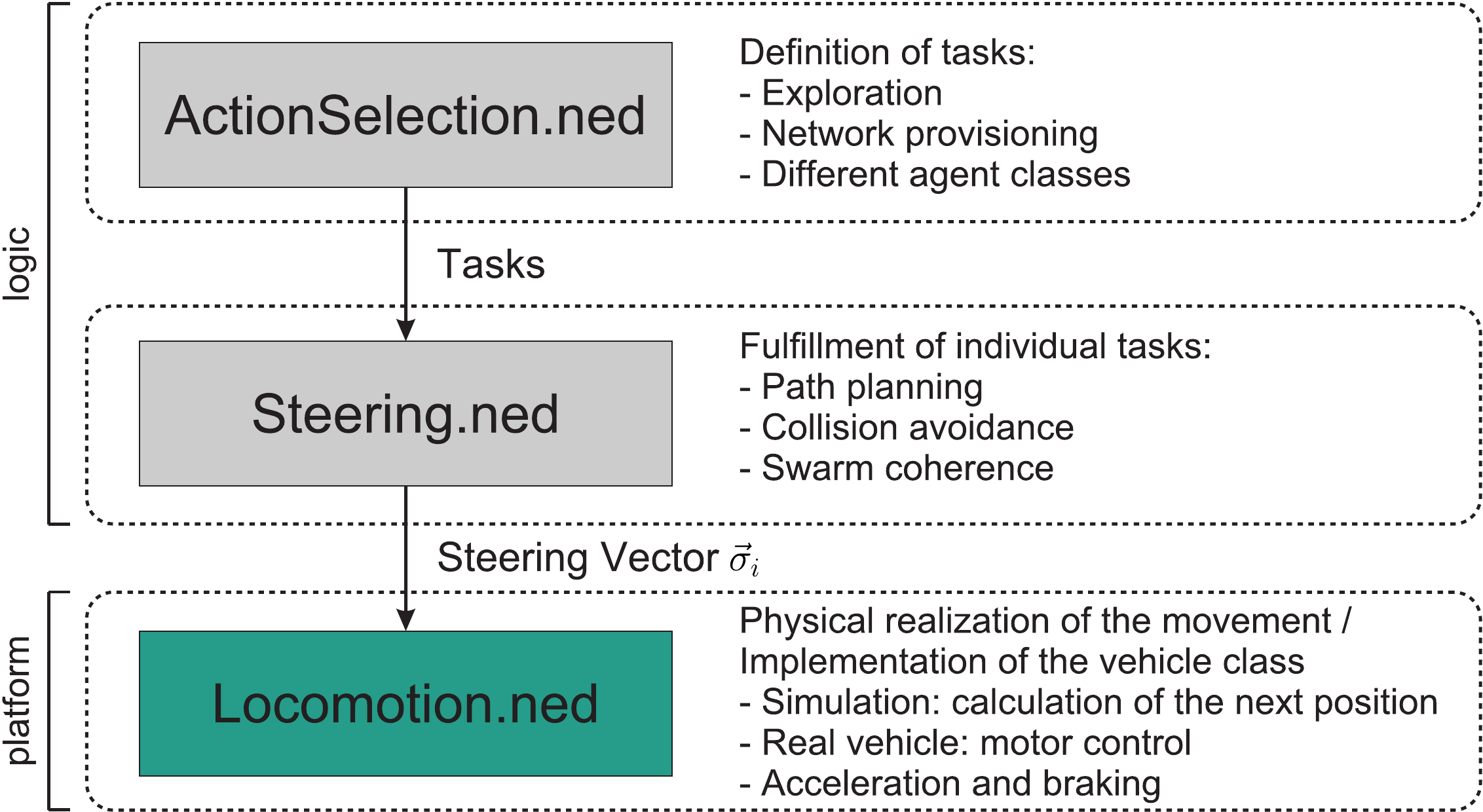}
	\caption{The Reynolds mobility model provides a meta-model for the integration of mobility algorithms and consists of multiple layers}
	\label{fig:reynolds}
	\vspace{-10pt}
\end{figure}

\begin{figure}[ht!]  	
	\centering		  
	\includegraphics[width=1\columnwidth]{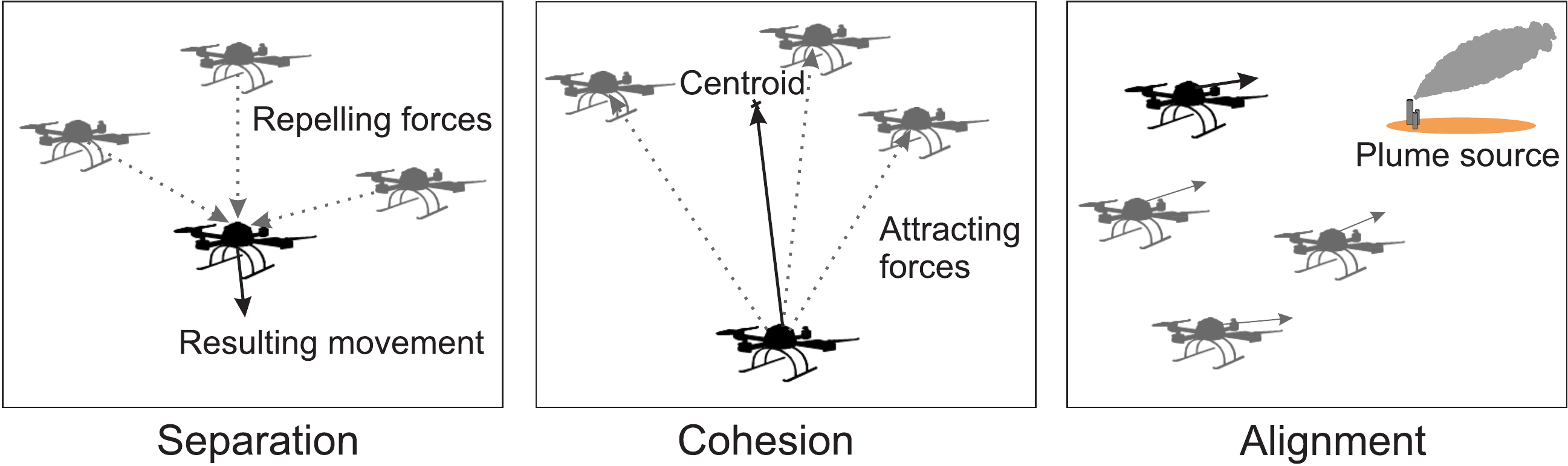}
	\caption{Reynolds' three basic rules for the behavior of individual agents travelling in swarms}
	\label{fig:swarm}
\end{figure}

In the field of \acp{MANET}, mobile hosts are often modelled using established generic mobility models. While this approach is sufficient for protocol evaluations of independent nodes, it is not capable of providing a realistic representation of the mobility behavior of cooperative autonomous agents. In real-world applications, mobile robots are influenced by the mobility characteristics of other nodes in order to avoid collisions, keep up communication paths and to fulfill cooperative tasks (e.g. exploration and network provisioning). Reynolds has proposed a basic mobility model for autonomous agents in \cite{Reynolds1999} that has received great acceptance inside the scientific community. We use this well-known approach as a general logical meta-model for our mobility modules. One of its key advantages is the intrinsic support for combining different behavior models, which can be used to increase the capabilities of situation-aware reacting. The model consists of three layers, which are illustrated in Fig.~\ref{fig:reynolds}.
\begin{itemize}
	\item \emph{Action Selection} defines the global tasks the mission has to fulfill. Examples are exploration and network provisioning. Different agent classes (e.g., scouts and relays) can be defined in order to pay attention to specific requirements of these tasks.
	\item \emph{Steering} determines how a specific task is handled. Usually multiple specific steerings are handled simultaneously to take the various challenges into account. 
	\item \emph{Locomotion} describes the physical movement of the nodes. For simulations a model of the vehicles' movement characteristics is used, while real-life vehicles control the actual motors.
\end{itemize}
The management of the interaction between the layers is performed by the \emph{ReynoldsMobilityModel.ned} module. It inherits from the \emph{MovingMobilityBase.ned} module and uses the parent module's update timer. With each mobility update all steerings are triggered and the steering vector $\vec{\sigma}_{i}$ is computed as their weighted mean. Different weights are applied to specify the importance of individual steerings, e.g., collision avoidance might be of higher priority than exploration if the distance between vehicles is low. The result vector is then passed to the \emph{Locomotion.ned} module, which maps the desired movement to a travelled path with respect to the physical capabilities of the vehicle. If different vehicle types (e.g., multicopters and planes) are modelled using dedicated locomotion classes, they can use the same steerings, increasing the portability of mobility algorithms.
Many application scenarios for mobile robotic networks (e.g., exploration of hazardous areas) focus on cooperative task fullfilment instead of individual mobility. In \cite{Reynolds1987}, Reynolds describes three basic rules for the behavior of individual agents when travelling in swarms, which are illustrated in Fig.~\ref{fig:swarm}.
\begin{itemize}
	\item\emph{Separation} is used to keep up a minimal distance between individual agents. In the field of \acp{UAV} this aspect represents the \emph{collision avoidance} mechanism. Recent implementations often make use of potential fields \cite{RuchtiSenkbeilCarrollEtAl2014}, where obstacles and other agents cause repelling forces.
	\item\emph{Cohesion} is the antagonist of separation and describes the maintenance of the swarm structure, which has a high impact on the number of available communication links. This aspect is important if network provisioning is the main task and it is often represented using potential fields with attracting forces to the swarm centroid \cite{GoddemeierRohdePojdaEtAl2011}.
	\item\emph{Alignment} describes the focus on a swarm target and addresses the fulfillment of cooperative tasks such as plume exploration \cite{BehnkeBoekWietfeld2013}.
\end{itemize}
Decent models of cooperative swarms have to address all of these aspects. Individual behavior algorithms can be implemented as steerings, which are then triggered by the meta-model.
Knowledge about the positions of other swarm participants is one of the key requirements for the design of cooperative mobility algorithms. This information is managed individually by each agent using the \emph{LocationService.ned} module, which acts as a mediator between the mobility control and the routing layer. After each position change, the own position entry in the location service is refreshed. Mobility information contained in routing messages is used to update the current knowledge about the sender node in the location service proving a solid data base for path planning. The data entries contain different types of individual mobility information.
\begin{itemize}
	\item The \emph{position} $\vec{P}_{i}$ represents the measured position value in the mobility update iteration $i$ and it is derived from the real position used by the \emph{MovingMobilityBase.ned} module. In order to take the influences of real-world \ac{GNSS} receiver inaccuracies into account it can be distorted with a noise vector for a defined maximum positioning error. Further extensions could contain different positioning methods like using the characteristics of wireless signals for location estimations.
	\item The \emph{steering vector} $\vec{\sigma}_{i}$ is the weighted superposition of individual steering results and represents the current movement vector to the desired position in the next iteration $i+1$.
	\item \emph{Waypoints} represent the planned trajectory of the agent. One example for obtaining this information in future real-world applications is the navigation system of autonomous cars.
\end{itemize}
Apart from modelling mobility behavior as steerings, which are executed in OMNeT++, mobility data can also be loaded from trace files. With this approach mobility data obtained from real-world measurements as well as from other simulators can directly be used in order to analyze the routing characteristics of the network.
With the availability of mobility data managed by the location service, nodes are able to predict future positions for a defined prediction width $N_{p}$ with the help of the \emph{TrajectoryPrediction.ned} module. This information can be used by the routing layer in order to select the path with the highest assumed stability to avoid packet losses. $N_{p}$ is a multiple of the mobility update interval $\Delta t_{u}$ and represents the tradeoff between the influence of the current and the future link-state for the total link quality assessment.

\subsection{Mobility-aware routing}

\begin{figure}[ht!]  	
	\centering		  
	\includegraphics[width=1\columnwidth]{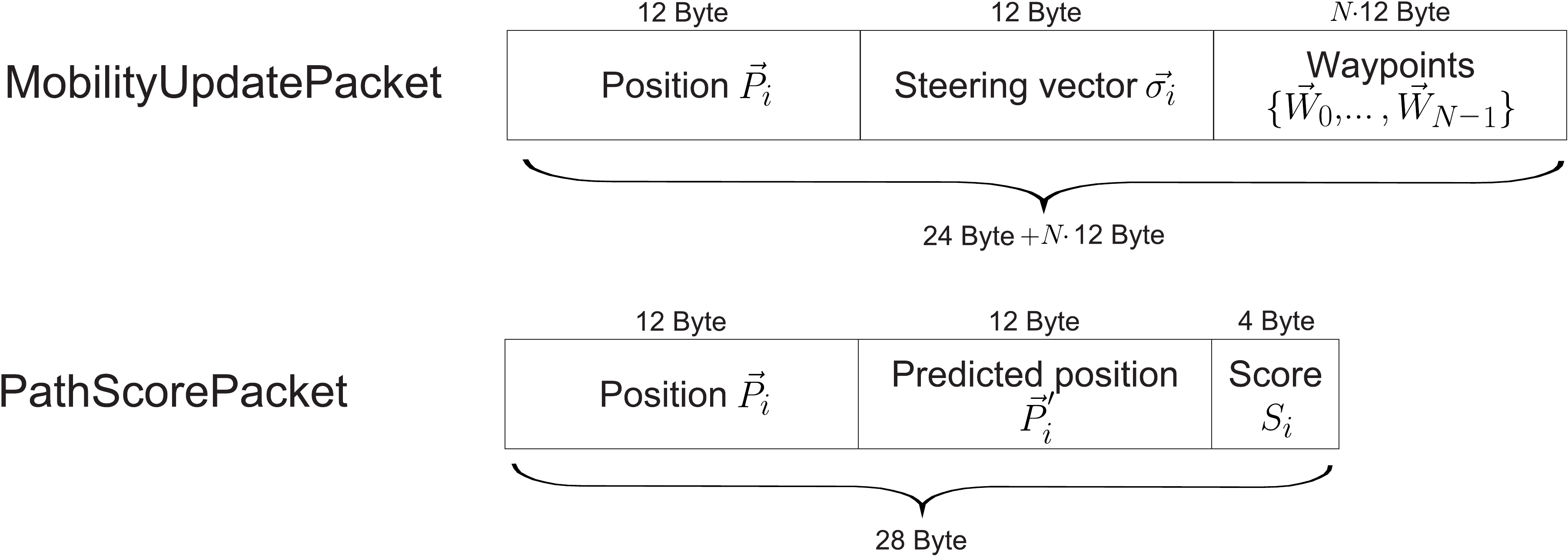}
	\caption{Message types for distributing the required mobility information through the network. The \emph{MobilityUpdatePacket} contains the individual mobility information needed for predictive path planning, while the \emph{PathScorePacket} only contains the prediction results needed for the stigmergic approach.}
	\label{fig:frames}
	\vspace{-20pt}	
\end{figure}

The selection of forwarder nodes for a given destination is the main task of the routing process and therefore a key factor for providing reliable communication. The mobility-aware approach uses current mobility data as well as predicted knowledge to improve the robustness of communication paths.
There are two general classes for using predicted mobility data for enhancing the routing decisions. \emph{Predictive path planning} requires all nodes to propagate their current mobility information through the network. Every node maintains a database of the mobility information of all others nodes, which is used to predict the future state of the whole network. The prediction is done with the trajectory prediction algorithm of \cite{SliwaBehnkeIdeEtAl2016}. Alg.~\ref{alg:dijkstra} describes the selection of the best forwarder node for a given destination. It assumes the routing protocol maintains topology information in a link map. Future links are assumed to be valid if their predicted distance $d$ is below a maximum distance $d_{max}$, which is obtained from a channel model for a defined	receiver sensitivity $P_{e,min}$. This channel model represents the assumption of the channel conditions from the agent's view and it is not equal to the one used by OMNeT++ to calculate the path loss. The actual distribution of mobility information is performed with the \emph{MobilityUpdatePacket} (see Fig.~\ref{fig:frames}) and can either be broadcasted or piggybacked to existing routing messages.
If mobility update packets are lost, the routing process may use outdated information.
\vspace{-10pt}
\begin{algorithm}[h!]  
\caption{Predictive Geo-based Dijkstra}\label{alg:dijkstra}
\begin{algorithmic}[1]
	\Function{findBestNeighbor}{$destination$, $linkMap$, $channelModel$, $dijkstra$, $P_{e,min}$}
	\State $bestNeighbor \gets null$
	\State $links \gets \lbrace \rbrace$
	\State $d_{max}\gets channelModel.calculateDistance(P_{e,min})$
	\For{$link$ in $linkMap$}\Comment{link is currently available}
		\State $node_{1}\gets link.firstNode$
		\State $node_{2}\gets link.secondNode$
		\State $d\gets calculatePredictedDistance(node_{1}, node_{2})$
		\If{$d<d_{max}$} \Comment{link is available in the future}
			\State{$links.append(link$)}
		\EndIf		
	\EndFor	
	\State $path \gets dijkstra.findPath(destination, links)$
		\If{$path.length>0$}
			\State{$bestNeighbor \gets path.first$}
		\EndIf
		\State \textbf{return} $bestNeighbor$
	\EndFunction
	\end{algorithmic}

\end{algorithm}
\vspace{-10pt}

With the \emph{stigmergic} approach used by B.A.T.Mobile, all agents periodically flood \emph{PathScorePackets} through the network, which are updated by each forwarder with the current and the predicted forwarder position and a score $S_{i}$ about the path quality of the reverse path to the sender. The value for $S_{i}$ is calculated with the \emph{path score metric}, which takes the absolute distance and the predicted distance development to the forwarder into account. 
Future implementations of routing protocols can inherit from the abstract \mbox{\emph{MobilityAwareRoutingBase.ned}} module, which provides access to current and predicted mobility data as well as the proposed routing metrics.
%
%
%

\section{Simulation-based system model}

\begin{figure}[ht!]  
	\vspace{-5pt}		
	\centering		  
	\includegraphics[width=1\columnwidth]{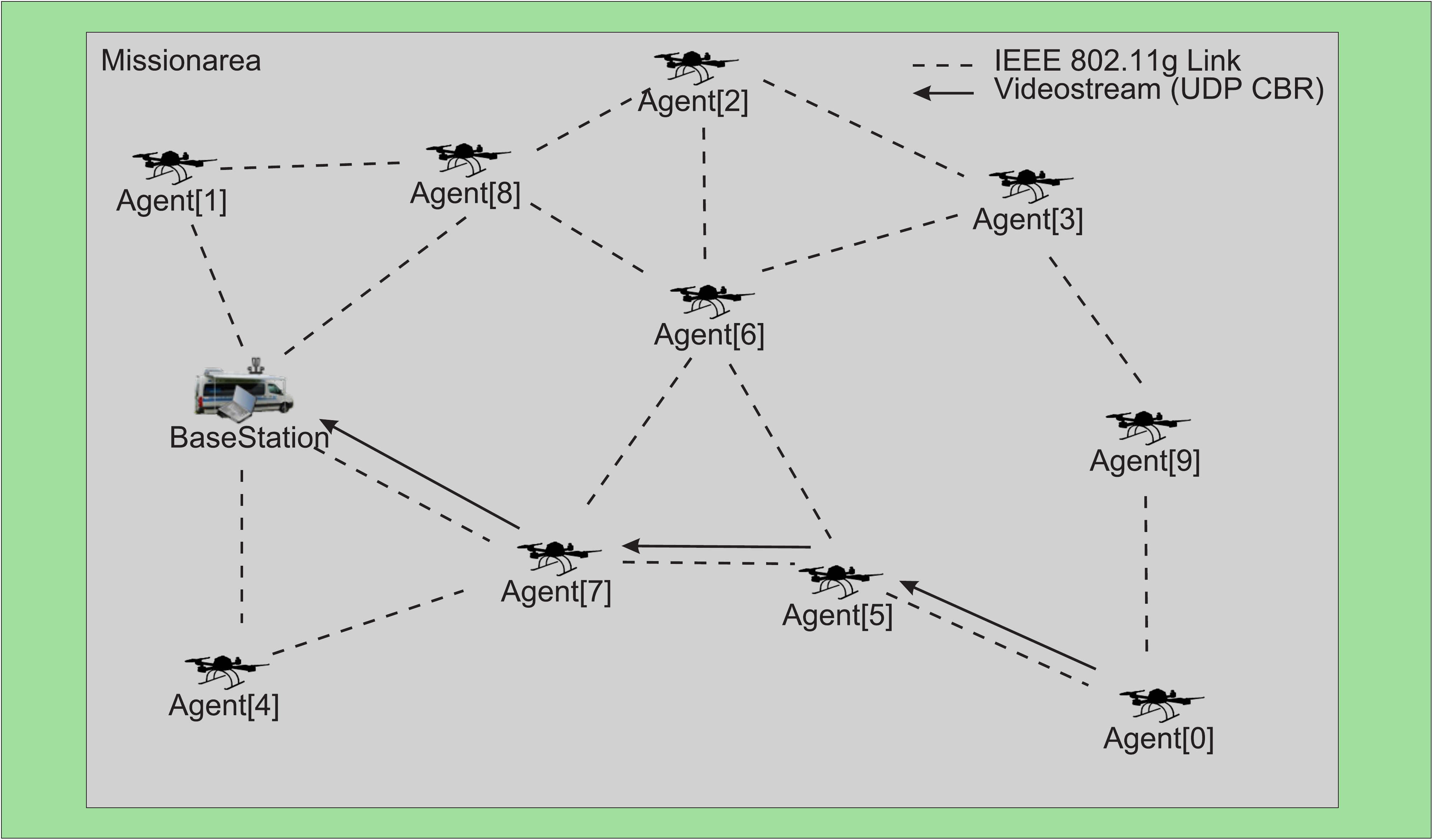}
	\caption{Simulation of the reference scenario in OMNeT++}
	\label{fig:net}
	\vspace{-10pt}	
\end{figure}

To provide an evaluation scenario, we consider a swarm of autonomous agents performing an exploration task in a defined mission area. For our reference scenario, we evaluate the air-to-ground \ac{PDR} of a videostream, which is transmitted from a randomly selected agent to a static base station as \ac{UDP} \ac{CBR} data. Fig~\ref{fig:net} illustrates the scenario.
A \emph{Controlled waypoint} mobility model is used for the exploration task in order to provoke frequent changes of the network topology during runtime. The waypoints are determined randomly but in contrast to the well-known \emph{Random waypoint} model, all nodes know about their future waypoints, making this information utilizable for the routing process. Additionally a collision avoidance mechanism is used and triggered if the distance of two nodes is below a defined threshold in order to avoid dangerous situations. Both mobility algorithms are implemented as \emph{steerings} and handled by the meta-modell.
The influence of packet losses is evaluated considering two different channel models. We use a \emph{Friis} channel model for rural environments and a \emph{Nakagami} (m=2) model for urban scenarios in order to identify implications of different packet loss probabilities for the routing mechanisms. The mobility-aware protocols \ac{MA-OLSR} and B.A.T.Mobile are compared to their base versions OLSR and B.A.T.M.A.N. The goal of the evaluation is a comparison of the  approaches predictive path planning and stigmergy in order to identify the one which is able to benefit most from the integration of mobility information and is a promising candidate for further extensions and evaluations.
Tab.~\ref{tab:simulation_parameters} shows the reference parameter assignment.
\vspace{-10pt}
\renewcommand{\arraystretch}{1.1}
\begin{table}[h]
	\centering
	\caption{Simulation Parameters for the reference scenario in OMNeT++/INETMANET}
	\begin{tabular}{|l|l|}
		\hline
		{\bf Simulation parameter} 			& {\bf Value}                     		  \\ \hline
		
		Mission area & 500 m x 500 m x 250 m  \\ \hline
		Number of agents & 10 \\ \hline

		Steering[0] (Exploration)& Controlled Waypoint \\ \hline
		Steering[0].weight & 1 \\ \hline	
		Steering[1] (Collision Avoidance) & LocationBasedCollAvoidance \\ \hline
		Steering[1].weight & 10 \\ \hline
		Steering[1].minDistance  & 30 m \\ \hline	
		Locomotion & UAVLocomotion \\ \hline
		
		Mobility update interval $\Delta t_{u}$ & 250 ms\\ \hline
		Movement speed & 50 km/h \\ \hline
		Channel model & Friis / Nakagami ($\alpha = 2.75$) \\ \hline
		Videostream bitrate & 2 Mbit/s \\ \hline
		Videostream packet size & 1460 Byte \\ \hline
		Telemetry broadcast interval & 250 ms  \\ \hline
		Telemetry packet size & 1000 Byte \\ \hline
		
		OGM broadcast interval (B.A.T.M.A.N.) & 0.5 s \\ \hline
		HELLO interval (OLSR) & 0.5 s \\ \hline
		\ac{TC} interval (OLSR) & 1 s \\ \hline
		
		\ac{MAC} layer & IEEE802.11g \\ \hline
		Bitrate & 54 Mbit/s \\ \hline
		Transport layer protocol & UDP \\ \hline
		
		Transmission power & 100 mW \\ \hline
		Carrier frequency & 2.4 GHz \\ \hline
		Receiver sensitivity ($P_{e,min}$)& -83 dBm \\ \hline
		
		Simulation time per run & 300 s \\ \hline
		Number of simulation runs & 50 \\ \hline
		
		Mobility data history size	$N_{h}$ & 5 \\ \hline
		Prediction width $N_{p}$ & 15 \\ \hline
	\end{tabular}
	\label{tab:simulation_parameters}
\vspace{-10pt}
\end{table}

\section{Proof of concept evaluation}

\begin{figure}[ht!]  	
	\centering		  
	\includegraphics[width=1\columnwidth]{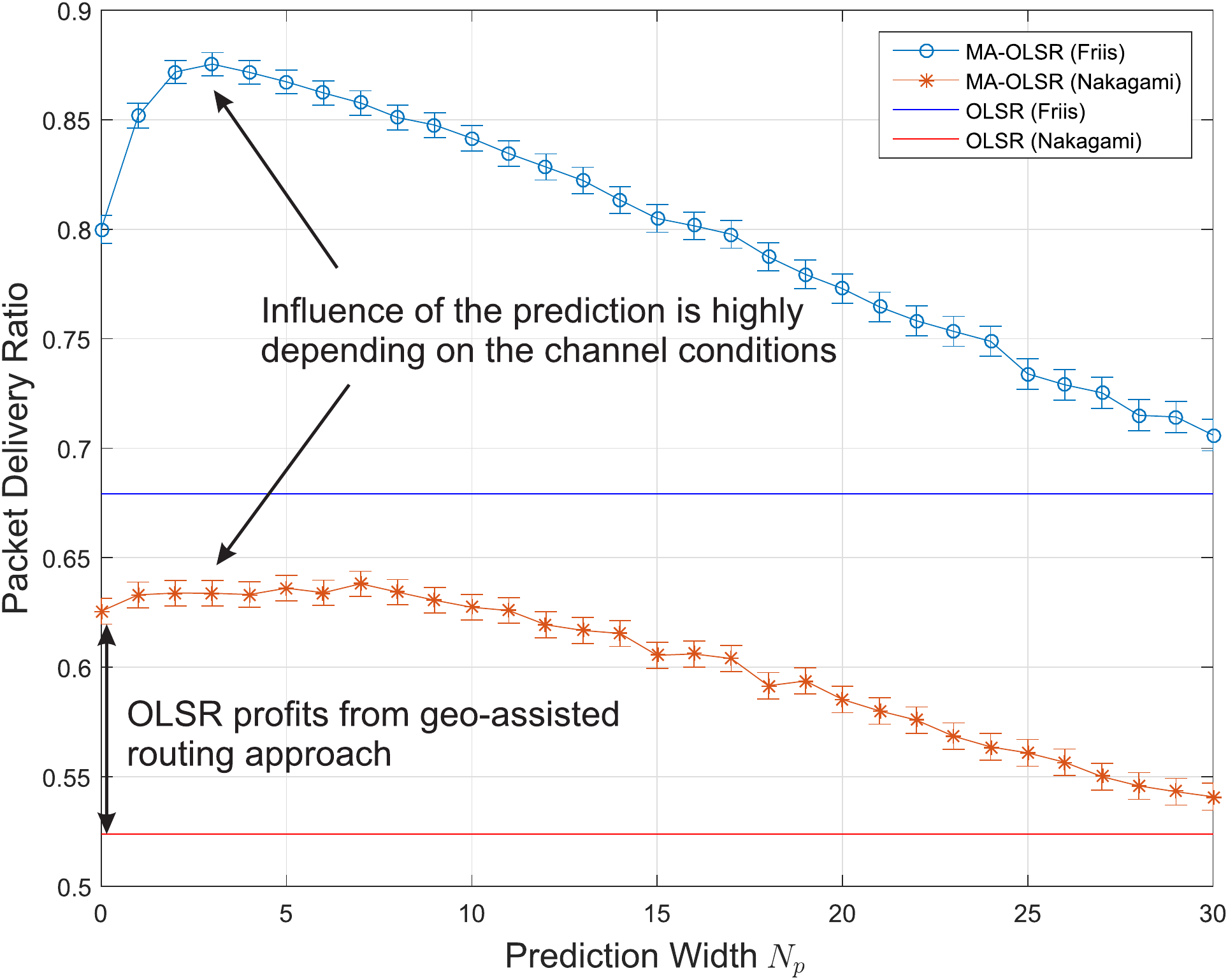}
	\caption{Performance evaluation of OLSR using predictive path planning}
	\label{fig:olsr}
\end{figure}

\begin{figure}[bh!]  
	\vspace{-5pt}	
	\centering		  
	\includegraphics[width=1\columnwidth]{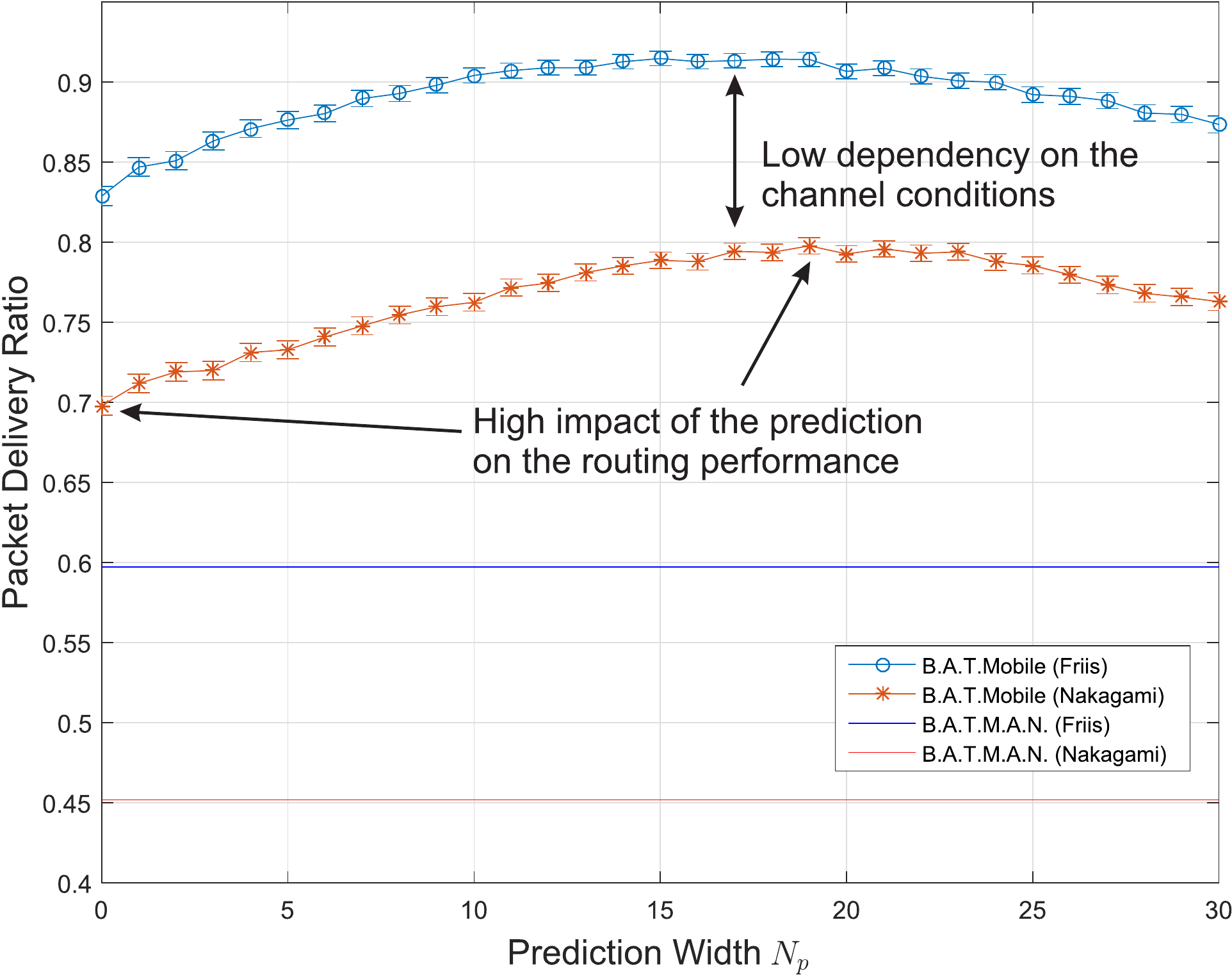}
	\caption{Performance evaluation of B.A.T.M.A.N. using predictive stigmergy}
	\label{fig:batman}
	\vspace{-10pt}
\end{figure}

In this section, we present example results obtained with the extension to INETMANET in order to prove its capability of providing a base for implementing further mobility-aware routing protocols and to show the efficiency of the described principles. The results show the 0.95 confidence interval of 50 simulation runs. 
Fig.~\ref{fig:olsr} compares the PDR values for OLSR and MA-OLSR, which uses the presented
\emph{predictive path planning} algorithm in two different channel models. 
%
%
MA-OLSR benefits from using geo-information in both scenarios. However, the impact of using predicted information is highly depending on the channel conditions. The path planning approach requires access to recent mobility data of all nodes in network. Packet losses can lead to serious prediction errors and have a negative effect on the routing performance.
The results of using the stigmergic prediction approach are shown in Fig.~\ref{fig:batman}. In contrast to MA-OLSR, B.A.T.Mobile achieves a high robustness against packet losses. As each packet contains information about its whole travelled path, the routing process does not necessarily require data from all nodes of the network to make a decision. These properties demonstrate the protocol as well as the routing metric in general are highly benefitting from the interaction with the mobility control layer. 

\section{Conclusion}

In this paper, we presented an extension to the INETMANET framework, which focuses on mobility-aware routing and uses a cross-layer approach to utilize mobility control information to enhance the routing process. The mobility behavior is implemented according to a well-known mobility meta-model that is able to provide a much more realistic representation of the mobility behavior of autonomous agents than generic algorithms. A new base module for mobility-aware protocols provides access to current and predicted mobility data as well as different routing metrics and information distribution methods. In a proof of concept evaluation, we demonstrated its potential to enhance the communication robustness even under challenging channel conditions. Overall, the consideration of mobility data for routing decision has shown to be a worthwile topic for further research and development. After the introduction of mobility-aware routing protocols, we want to extend the capabilities of the framework with communication-aware mobility algorithms. Furthermore, we want to add support for further vehicle classes. In this context the application and evaluation of the proposed routing protocols in the field of \acp{VANET} is another promising research topic.

\section*{Acknowledgment}
\vspace{-3pt}
\footnotesize
Part of the work on this paper has been supported by Deutsche Forschungsgemeinschaft (DFG) within the Collaborative Research Center SFB 876 ``Providing Information by Resource-Constrained Analysis'', project B4.

\bibliographystyle{IEEEtran}
\bibliography{MobilityAwareRoutingPaper}

\end{document}